\documentclass[12pt]{article}

\usepackage{latexsym} 
\usepackage{amssymb}  
\usepackage{epsfig}       

\newcommand{\txt}{\textstyle}

\newcommand{\beq}{\begin{equation}}
\newcommand{\eeq}{\end{equation}}
\newcommand{\ba}{\begin{array}}
\newcommand{\bea}{\begin{eqnarray}}
\newcommand{\ea}{\end{array}}
\newcommand{\eea}{\end{eqnarray}}

\newcommand\comment[1]{ \hbox{[{\it Comment suppressed here.}\/]} }
\newcommand\hide[1]{}

\newcommand{\tr}{\hbox{tr}}
\newcommand{\Tr}{\hbox{Tr}}

\newcommand{\bx}{{\bf x}}

\newcommand{\skipover}[1]{}

\newcommand{\C}{{\cal C}}
\newcommand{\half} {{\txt {1\over 2}}}

\newcommand{\bra}{\langle}
\newcommand{\ket}{\rangle}

\newcommand{\vecz}{{\mathbf z}}

\makeatletter 

\def\appendix{\par                              
    \setcounter{section}{0}                     
    \setcounter{subsection}{0}
    \renewcommand{\theequation}{\Alph{section}.\arabic{equation}}
    \renewcommand{\thesection}{Appendix \Alph{section}
                \setcounter{equation}{0}  } 
}

\def\applabel#1{\@bsphack
  \protected@write\@auxout{}%
         {\string\newlabel{#1}{{\Alph{section}}{\thepage}}}%
  \@esphack}

\def\section{
\setcounter{equation}{0}        
\@startsection {section}{1}{\z@}{-3.5ex plus -1ex minus 
 -.2ex}{2.3ex plus .2ex}{\large\bf}}
\renewcommand{\theequation}{\arabic{section}.\arabic{equation}}

\def\subsection{\@startsection{subsection}{2}{\z@}{-3.25ex plus -1ex minus 
 -.2ex}{1.5ex plus .2ex}{\normalsize\bf}}

\def\subsubsection{\@startsection{subsubsection}{3}{\z@}{-3.25ex plus
 -1ex minus -.2ex}{1.5ex plus .2ex}{\normalsize}}

\makeatother   

\newsavebox{\eqlabel}

\makeatletter  
\newlength{\numblen}
\newsavebox{\eqnumb}
\def\@eqnnum{\savebox{\eqnumb}{\rm (\theequation)}%
\settowidth{\numblen}{\usebox{\eqnumb}}%
\makebox[\numblen][l]{\usebox{\eqnumb}~~~\usebox{\eqlabel}}}
\makeatother   

\newenvironment{equationwithlabel}[1]{ %
  \begin{equation}\label{#1} }{\end{equation}} 
\newcommand{\beql}[1]{\begin{equationwithlabel}{#1}}
\newcommand{\eeql}{\end{equationwithlabel}}
\newenvironment{equationarraywithlabel}[1]{ %
  \begin{eqnarray}\label{#1} }{\end{eqnarray}} 
\newcommand{\beal}[1]{\begin{equationarraywithlabel}{#1}}
\newcommand{\eeal}{\end{equationarraywithlabel}}

\begin{document}

\title{
\vskip -100pt
{
\begin{normalsize}
\mbox{} \hfill hep-ph/0201308
\vskip  70pt
\end{normalsize}
}
Far-from-equilibrium dynamics with broken symmetries 
from the $2PI$--$1/N$ expansion}

\author{
Gert Aarts\thanks{email: aarts@mps.ohio-state.edu} $^a$,
\addtocounter{footnote}{1}
Daria Ahrensmeier\thanks{email: dahrens@physik.uni-bielefeld.de} $^b$,  
Rudolf Baier\thanks{email: baier@physik.uni-bielefeld.de} $^b$,\\
\addtocounter{footnote}{1}
J\"urgen Berges\thanks{email: j.berges@thphys.uni-heidelberg.de} $^c$
and
Julien Serreau\thanks{email: serreau@thphys.uni-heidelberg.de} $\,\,^c$
\\
[2.ex]
\normalsize{$^a$ Department of Physics, The Ohio State University}\\
\normalsize{174 West 18th Avenue, Columbus, OH 43210, USA}\\
[1.ex]
\normalsize{$^b$ Fakult{\"a}t f{\"ur} Physik, Universit{\"a}t Bielefeld}\\
\normalsize{Universit{\"a}tsstra{\ss}e, 33615 Bielefeld, Germany}\\
[1.ex]\normalsize{$^c$ Universit{\"a}t Heidelberg, Institut f{\"u}r 
Theoretische Physik}\\
\normalsize{Philosophenweg 16, 69120 Heidelberg, Germany}
}

\date{\normalsize{January 31, 2002 $\,\,$}}

\begin{titlepage}
\maketitle
\def\thepage{}          

\begin{abstract}

We derive the nonequilibrium real-time evolution of an $O(N)$ -- invariant
scalar quantum field theory in the presence of a nonvanishing expectation
value of the quantum field. Using a systematic $1/N$ expansion of the
$2PI$ effective action to next-to-leading order, we obtain nonperturbative
evolution equations which include scattering and memory effects. The
equivalence of the direct method, which requires the resummation of an
infinite number of skeleton diagrams, with the auxiliary-field formalism,
which involves only one diagram at next-to-leading order, is shown.

\end{abstract}

\end{titlepage}

\renewcommand{\thepage}{\arabic{page}}


\section{Introduction}
\label{sectionintro}

Nonequilibrium quantum field theory has a wide range of applications,
including current and upcoming relativistic heavy-ion collision
experiments at RHIC and LHC, phase transitions in the early universe
or the formation of Bose-Einstein condensates in the laboratory. Important
theoretical progress has been achieved with effective descriptions based
on a separation of scales in the weak-coupling limit
\cite{Bodeker:2001pa}, or for systems close to equilibrium using
approximations such as (non)linear response or gradient expansions
\cite{proceeding}. However, the description of far-from-equilibrium
dynamics is still in its infancies. The situation is complicated by the
fact that typically there is no clear separation of scales which is valid
at all times and it is often difficult to identify a small expansion
parameter. For example, standard perturbation theory is plagued by the
problem that a secular (unbounded) time evolution prevents the description
of the late-time behavior of quantum fields.

Practicable nonperturbative approximation schemes may be based on the
two-particle irreducible ($2PI$) generating functional for Green's
functions \cite{Cornwall:1974vz}. Recently, a systematic $1/N$ expansion
of the $2PI$ effective action has been proposed and applied to a scalar
$O(N)$--symmetric quantum field theory \cite{Berges:2001fi}. The approach
provides a controlled nonperturbative description of far-from-equilibrium
dynamics at early times as well as the late-time approach to thermal
equilibrium and can be applied in extreme nonequilibrium situations
\cite{Berges:2001fi,Aarts:2001yn}. This is in contrast to the standard
$1/N$ expansion of the $1PI$ effective action which is secular in time once
direct scattering is taken into account \cite{Bettencourt:1997nf}. The
$2PI$--$1/N$ expansion extends previous successful descriptions of the
large-time behavior of quantum fields \cite{Berges:2000ur,Aarts:2001qa},
which employ the loop expansion of the $2PI$ effective action relevant at
weak couplings \cite{Cornwall:1974vz,KadanoffBaym,Calzetta:1986cq}.  For
systems in or close to equilibrium recent applications of the loop
expansion can be found in Refs.\
\cite{Ivanov:1998nv,vanHees:2002js,Blaizot:1999ip,Braaten:2001en}. A $1/N$
expansion has the advantage over the loop expansion that it is not
restricted to small couplings, an observation made in the context of
nonequilibrium quantum field dynamics already for quite some time
\cite{LOapp}. However, in order to describe quantum scattering and
thermalization the inclusion of the usually discarded next-to-leading
order (NLO) contributions is crucial.

In Ref.\ \cite{Berges:2001fi} the $2PI$--$1/N$ expansion has been carried
out to NLO in the symmetric regime for a vanishing expectation value
$\phi$ of the quantum field. Here, we derive the evolution equations from
the NLO approximation of the $2PI$ effective action for nonzero $\phi$. A
nonvanishing field expectation value is important to describe the physics
of heavy ion collisions, where the presence of a substantial scalar
quark--antiquark condensate signals the spontaneous breakdown of chiral
symmetry. (In the language of the $O(4)$ linear sigma model for
two-flavor QCD one has $\phi \sim \langle \bar{q} q \rangle$.) Important
nonequilibrium applications in this context include the formation of
disoriented chiral condensates (DCCs, cf.\ \cite{DCC}) or the decay
\cite{Ahrensmeier:2000pg} of parity odd metastable states in hot QCD
\cite{Kharzeev:1998kz}. Similar far-from-equilibrium applications in
inflationary cosmology concern the phenomenon of preheating, where the
dynamics of the inflaton $\phi$ is expected to trigger explosive particle
production (see, e.g., \cite{preheat}). The availability of a quantum
field description of the dynamics is important for cases where the
effectively classical approximation (in the context of inflation, see
e.g.\ \cite{Khlebnikov:1996mc}) may not be appropriate.

In this work, we perform a systematic $1/N$ expansion of the $2PI$
effective action to next-to-leading order. A detailed discussion of the
classification of diagrams is given in \mbox{Sec.\ \ref{Nexpansion}}.  This
allows us to give the effective action (Sec.\ \ref{2PINLOaction}) and the
equations of motion (Sec.\ \ref{Secteom}) in a straightforward manner. In
\mbox{Sec.\ \ref{auxiliary}} it is shown how this result is obtained using the
auxiliary-field formalism \cite{Mihaila:2000sr}. The causal equations for
the spectral and statistical functions are listed in Sec.\
\ref{secevolution} and a perturbative approximation to second order in
the coupling constant is given in Sec.\ \ref{weakcoupling}. In two
Appendices we discuss further features of the equations and the
realization of Goldstone's theorem.

\section{The $2PI$ effective action}
\label{section2pi}

We consider a real scalar $N$--component quantum field 
$\varphi_a$ ($a=1,\ldots, N$) with a classical $O(N)$--invariant action 
\beql{classical}
S[\varphi] = \int d^{d+1}x \left[ \frac{1}{2} 
\partial_{x^0} \varphi_a
\partial_{x^0} \varphi_a
-\frac{1}{2} \partial_{\bx} \varphi_a
\partial_{\bx} \varphi_a
- \frac{1}{2} m^2 \varphi_a \varphi_a
- \frac{\lambda}{4! N} \left(\varphi_a\varphi_a\right)^2 \right].
\eeql
Summation over repeated indices is implied and $x \equiv (x^0,\bx)$. 
All correlation functions of the quantum theory can be obtained from the 
effective action $\Gamma[\phi,G]$, the two-particle irreducible ($2PI$)
generating functional for  Green's functions parametrized 
by the macroscopic field $\phi_a(x)$ 
and the composite field $G_{ab}(x,y)$, given by
\begin{eqnarray}
\phi_a(x) & = & \langle\varphi_a(x)\rangle,\\
G_{ab}(x,y) & = & \langle T_{\cal C}\varphi_a(x)\varphi_b(y)\rangle
  -\langle\varphi_a(x)\rangle\langle\varphi_b(y)\rangle.
\end{eqnarray}
The brackets denote the expectation value with respect to the density matrix 
and $T_{\cal C}$ denotes time-ordering along a contour ${\cal C}$ in the
complex time plane. At this stage the explicit form of the contour
is not needed.

A discussion of the defining functional integral of the
$2PI$ effective action can be found in Ref.\ \cite{Cornwall:1974vz}.
Following \cite{Cornwall:1974vz} it is convenient to parametrize the $2PI$ 
effective action as
\beql{2PIaction}
\Gamma[\phi,G] = S[\phi] + \frac{i}{2} \Tr\ln G^{-1} 
          + \frac{i}{2} \Tr\, G_0^{-1}(\phi)\, G
          + \Gamma_2[\phi,G] +{\rm const}. 
\eeql 
Here the classical inverse propagator $i G_{0}^{-1}(\phi)$ 
is given by 
\bea
&&i G^{-1}_{0,ab}(x,y;\phi) \equiv 
  \frac{\delta^2 S[\phi]}{\delta\phi_a(x)\delta\phi_b(y)}\nonumber\\
\label{classprop}
&&    
= - \left(
\left[ \square_x + m^2 + \frac{\lambda}{6 N}\,\phi^2(x) \right]
      \delta_{ab} + \frac{\lambda}{3 N} \phi_a(x)\phi_b(x)
\right)\delta_{\C}(x-y)
\eea
with $\phi^2 \equiv \phi_a\phi_a$ and $\delta_{\C}(x-y) \equiv
\delta_{\C}(x^0-y^0) \delta^{d}({\bf x}-{\bf y})$.
The contribution $\Gamma_2[\phi,G]$ is given by all closed
$2PI$ graphs\footnote{A graph is two--particle irreducible 
if it does not become disconnected upon opening two propagator lines.} 
with the propagator lines set equal to $G$ 
\cite{Cornwall:1974vz}. The effective interaction vertices are
obtained from the terms cubic and higher in $\varphi$ in the classical 
action (\ref{classical}) after shifting the field $\varphi \rightarrow 
\phi + \varphi$. 
In presence of the nonvanishing expectation value $\phi$ this results 
in a cubic and a quartic vertex
\beql{interaction}
{\cal L}_{\rm int}(x;\phi,\varphi) =  
- \frac{\lambda}{6 N} \phi_a(x)\varphi_a(x)\varphi_b(x)\varphi_b(x)
- \frac{\lambda}{4! N} \left[\varphi_a(x)\varphi_a(x)\right]^2.
\eeql

Dynamical equations for $\phi_a$ and $G_{ab}$ can be found by minimizing 
the effective action. In the absence of external sources
physical solutions require 
\begin{equation}
\frac{\delta \Gamma[\phi,G]}{\delta \phi_a(x)} = 0, 
\label{phistationary}
\end{equation}
which leads to the macroscopic field evolution equation
\bea
 \nonumber
-\left(\square_x + m^2 + \frac{\lambda}{6N}
\left[ \phi^2(x) + G_{bb}(x,x) \right] \right) \phi_a(x)  &=& 
 \frac{\lambda}{3N} \phi_b(x)G_{ba}(x,x) \\ 
&& - \frac{\delta \Gamma_2[\phi,G]}{\delta \phi_a(x)},
\label{feeq}
\eea
as well as
\begin{equation}
\frac{\delta \Gamma[\phi,G]}{\delta G_{ab}(x,y)} = 0,
\label{stationary}
\end{equation}
which leads to  
\begin{equation}
G_{ab}^{-1}(x,y) = G_{0,ab}^{-1}(x,y) - \Sigma_{ab}(x,y;\phi,G)
\label{gap}
\end{equation}
with
\begin{equation}
\label{eqself}
\Sigma_{ab}(x,y;\phi,G) \equiv 
2i\frac{\delta \Gamma_2[\phi,G]}{\delta G_{ab}(x,y)}.
\end{equation}
Eq.\ (\ref{gap}) can be rewritten as a partial differential equation suitable
for initial value problems by convolution with $G$, 
\begin{equation} 
\int_z G_{0,ab}^{-1}(x,z)G_{bc}(z,y) = \int_z 
\Sigma_{ac}(x,z)G_{cb}(z,y) + \delta_{ab}\delta_\C(x-y),
\end{equation}
where the shorthand notation
$\int_z = \int_\C dz^0 \int d\vecz$ is employed. With the classical inverse 
propagator (\ref{classprop}) this differential equation reads explicitly
\bea
\nonumber
-\left[\square_x + m^2 + \frac{\lambda}{6 N}\phi^2(x) \right]
G_{ab}(x,y) = 
\frac{\lambda}{3 N} \phi_a(x)\phi_c(x)G_{cb}(x,y) &&\\
\label{geeq}
+i \int_z \Sigma_{ac}(x,z;\phi,G) G_{cb}(z,y) + i \delta_{ab} 
\delta_{\C}(x-y). &&
\eea
The evolution of $\phi_a$ and $G_{ab}$ is determined by Eqs.\ 
(\ref{feeq}) and (\ref{geeq}), once $\Gamma_2[\phi,G]$ is specified.

\section{The $2PI$--$1/N$ expansion}
\label{Nexpansion} 

In this section we discuss the $2PI$--$1/N$ expansion proposed in Ref.\
\cite{Berges:2001fi} in more detail. We present a simple classification
scheme based on $O(N)$--invariants which parametrize the $2PI$ diagrams
contributing to $\Gamma[\phi,G]$. The effective action at a given order in
$1/N$ can then be obtained from a straightforward summation of the
diagrams contributing at that order.  The NLO result for $\Gamma[\phi,G]$
for general field configurations $(\phi$, $G)$ is presented in 
\mbox{Sec.\ \ref{2PINLOaction}}. Below we will compare the direct
summation procedure described here with an alternative method, which
employs an auxiliary field to simplify the summation (cf.\ Sec.\
\ref{auxiliary}).
  
\subsection{Counting rules for irreducible $O(N)$-invariants} 

Following standard procedures \cite{Coleman:jh} the interaction term of
the classical action in Eq.\ (\ref{classical}) is written such that
$S[\phi]$ scales proportional to $N$. From the fields $\phi_a$ alone one
can construct only one independent invariant under $O(N)$ rotations, which
can be taken as $\tr\, \phi\phi \equiv \phi^2 = \phi_a \phi_a \sim
N$.  The minimum $\phi_0$ of the classical effective potential for this
theory is given by $\phi_0^2 = N (-6 m^2/\lambda)$ for negative
mass-squared $m^2$ and scales proportional to $N$.  Similarly, the trace
with respect to the field indices of the classical propagator $G_{0}$ is
of order~$N$.

The $2PI$ effective action is a singlet under $O(N)$ rotations and
parametrized by the two fields $\phi_a$ and $G_{ab}$.  To write down the
possible $O(N)$ invariants, which can be constructed from these fields, we
note that the number of $\phi$--fields has to be even in order to
construct an $O(N)$--singlet. For a compact notation we use 
$(\phi \phi )_{ab} = \phi_a \phi_b$. Consider a general singlet composed
of arbitrary powers of $(\phi\phi)_{ab}$ and $G_{ab}$ (we neglect all
space-time dependencies)
\bea
\nonumber
s &=& \tr \left[ (\phi \phi)^{p_1} \, G^{q_1} \, ... \, 
 (\phi \phi)^{p_n} \, G^{q_n} \right] \\
&=& (\phi^2 )^{p_1 + ... + p_n - n} \,
 \tr \left(\phi \phi \, G^{q_1}\right) \, ... \, 
 \tr \left(\phi \phi \, G^{q_n}\right)
\eea
for $p_i \not = 0$, where the $p_i$ and $q_i$ ($i=1,\ldots,n$)  are any
positive integer. (For all $p_i = 0$ the RHS is given by $\tr\,
G^{q_1+\ldots+q_n}$.)  The second line in the above equation follows from
simple contraction of the field indices and corresponds to the fact that
all functions of $\phi$ and $G$, which are singlets under $O(N)$, can be
built from the irreducible (i.e.\ nonfactorizable in field-index space)  
invariants
\beq
\phi^2, \quad\quad \tr (G^n) \quad\quad \mbox{and}  
\quad\quad \tr (\phi \phi G^n). 
\label{oninvariants}
\eeq
We note that for given $N$ only the invariants with $n \le N$ are
irreducible. However, we will see below that for lower orders in $1/N$ and
for sufficiently large $N$ one has $n < N$. In particular, for the
next-to-leading order approximation one finds that only invariants with $n
\le 2$ appear.

Since each single graph contributing to $\Gamma[\phi,G]$ is an
$O(N)$--singlet, we can express them with the help of the set of
invariants in Eq.\ (\ref{oninvariants}). The factors of $N$ in a given
graph have two origins: each irreducible invariant is taken to scale
proportional to $N$ since it contains exactly one trace over the field
indices, while each vertex provides a factor of $1/N$. The leading order
(LO) graphs then scale proportional to $N$ (as the classical action $S$),
next-to-leading order (NLO) contributions are of order one, the
next-to-next-to-leading order (NNLO) scales as $1/N$ and so on. This
provides a well-defined expansion of $\Gamma[\phi,G]$ in powers of $1/N$.  
We stress that by construction each order in the $2PI$--$1/N$ expansion
respects $O(N)$ symmetry.  In particular, this is crucial for the validity
of Goldstone's theorem in the case of spontaneous symmetry breaking.  For
constant $\phi \not = 0$ one observes that the mass matrix $\sim \delta^2
\Gamma[\phi,G^{\rm (stat)} (\phi)]/\delta\phi_a\delta \phi_b$, with
$G^{\rm (stat)} (\phi)$ being the solution of Eq.\ (\ref{stationary}), contains
$(N-1)$ massless ``transverse'' modes and one massive ``longitudinal''
mode (see Appendix A for a discussion).

\subsection{Classification of diagrams}

\begin{figure}[t]
\begin{center}
\epsfig{file=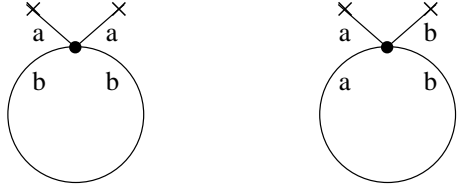,width=6.cm}
\end{center}
\caption{
Graphical representation of the $\phi$--dependent contributions for
$\Gamma_2 \equiv 0$. The crosses denote field insertions $\sim
\phi_a\phi_a$ for the left figure, which contributes at leading order, and
$\sim \phi_a\phi_b$ for the right figure contributing at next-to-leading
order.
}
\label{oneloopfig}
\end{figure}
In the following we will classify the various diagrams contributing to
$\Gamma[\phi,G]$. The expression (\ref{2PIaction}) for the $2PI$ effective
action contains, besides the classical action, the one-loop contribution
proportional to $\Tr\,\ln G^{-1} + \Tr\, G_0^{-1}(\phi) G$ and a
nonvanishing $\Gamma_2[\phi,G]$ if higher loops are taken into account.
The one-loop term contains both LO and NLO contributions.  The logarithmic
term corresponds, in absence of other terms, simply to the free field
effective action and scales proportional to the number of field
components $N$. To separate the LO and NLO contributions at the one-loop
level consider the second term $\Tr\, G_0^{-1}(\phi) G$. From the form of
the classical propagator (\ref{classprop}) one observes that it can be
decomposed into a term proportional to $\tr(G) \sim N$ and terms
(neglecting all the space-time structure)
\beq
\label{oneloop}
 \sim \frac{\lambda}{6N} \left[ \tr(\phi\phi)\,\tr(G) + 
 2\,\tr(\phi\phi G) \right].
\eeq
This can be seen as the sum of two ``$2PI$ one-loop graphs'' with field
insertion $\sim \phi_a\phi_a$ and $\sim \phi_a\phi_b$, respectively.
Counting the factors of $N$ coming from the traces and the prefactor, one
observes that only the first of the two terms in (\ref{oneloop})  
contributes at LO, while the second one is of order one.

\begin{figure}[t]
\begin{center}
\epsfig{file=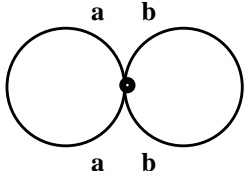,width=3.2cm}
\end{center}
\caption{LO contribution to the $2PI$ effective action.} 
\label{LOfig}
\end{figure}

\begin{figure}[b]
\begin{center}
\epsfig{file=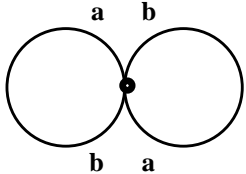,width=3.2cm}
\end{center}
\caption{NLO ``double bubble'' contribution.} 
\label{NLObubblefig}
\end{figure}

We now turn to $\Gamma_2[\phi,G]$ which contains all $2PI$ diagrams beyond
the one-loop level. The graphs are constructed from the three-point vertex
and the four-point vertex in the interaction Lagrangian
(\ref{interaction}) with $G$ associated to the propagator lines. From the
three-vertex $\sim \phi_a$ one easily observes that one cannot construct a
diagram which has a field insertion $\sim
\phi^2 = \phi_a\phi_a$. As a consequence, all loop diagrams beyond the
one-loop level can only depend on the basic invariants $\tr (G^n)$
and $\tr (\phi \phi \, G^n)$. Furthermore, we note that the invariant $\tr
(G) = G_{aa}$ can only appear if the graph contains a tadpole as shown in
Figs.\ \ref{LOfig} and \ref{NLObubblefig}.
The first graph, Fig.~\ref{LOfig}, is proportional to the product of the
term $[\tr(G)]^2$ and a factor $1/N$ from the quartic vertex. It therefore
contributes at LO to the effective action. The graph in 
Fig.~\ref{NLObubblefig} is proportional to $\tr(G^2)/N$ and of order one.  The
two-loop graphs shown in these figures are indeed the only two-particle
irreducible graphs which contain a tadpole. As a consequence, the only
invariants that can arise beyond two loops are $\tr (G^{n\geqslant2})$ and
$\tr (\phi \phi \, G^{n\geqslant1})$.

Using these considerations, one can now straightforwardly continue to
classify all the diagrams at a given order of the $1/N$ expansion.
Consider a graph with $V_3$ three-point and $V_4$ four-point vertices.  
The number of internal lines $I$ is determined by the familiar relation
\beq
 2I = 3V_3 + 4V_4.
\label{lines}
\eeq
We observe again that $V_3$ has to be even. This graph contains $V_3/2$
field insertions $\sim \phi\phi$, $2V_4 + \frac{3}{2} V_3$ propagators
$G$ and comes with a factor $(\lambda/N)^{V_3+V_4}$ from the vertices. The
highest power of $N$ is obtained by contracting the $\phi$ and the $G$
field such that the largest number of invariants $\tr (G^{n\geqslant2})$
and $\tr (\phi \phi \, G^{n\geqslant1})$ appears. The structure with the
highest power of $N$,
\beq
 \sim \Big[ \frac{\lambda}{N} \Big]^{V_3 + V_4}  
 \Big[ \tr( \phi \phi G) \Big]^{\frac{V_3}{2}}  
 \Big[ \tr( G^2 ) \Big]^{V_4+\frac{V_3}{2}}, 
\eeq
is of order one and contributes therefore at NLO. 

\begin{figure}
\begin{center}
\epsfig{file=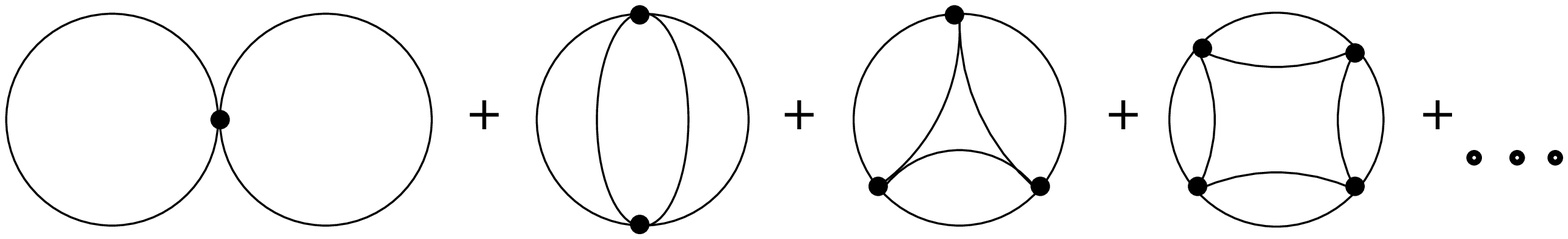,width=12.cm}
\end{center}
\caption{
NLO $\phi$--independent contribution to the $2PI$ effective action. Higher
loop diagrams in the infinite series can be obtained from the previous one
by introducing another ``rung'' with two propagator lines at each vertex.
The resummed series including the prefactors not displayed in the figure
is given by the first term in Eq.\ (\ref{NLOcont}).
} 
\label{NLOfig}
\end{figure}

\begin{figure}
\begin{center}
\epsfig{file=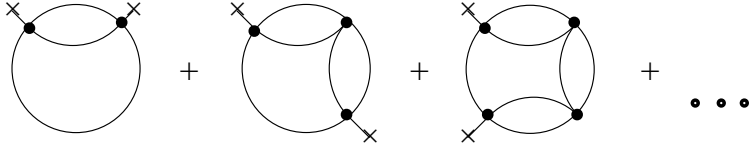,width=10.0cm}
\end{center}
\caption{
NLO $\phi$--dependent contribution to the $2PI$ effective action. Each
diagram in the infinite series can be obtained from the previous one by
introducing another ``rung'' with two propagator lines at each vertex.  
The resummed series is given by the second term in Eq.\ (\ref{NLOcont}).
The complete NLO contribution is given by the sum of the diagrams in 
Figs.\ \ref{NLOfig} and \ref{NLOfigb}.
} 
\label{NLOfigb}
\end{figure}

For $V_3=0$ the corresponding diagrams are shown in Fig.\ \ref{NLOfig}
\cite{Berges:2001fi}. For $V_3=2$ the infinite series of graphs is
presented in Fig.\ \ref{NLOfigb}.  We will now argue that graphs with
$V_3\geq 4$ are not two-particle irreducible such that we have classified  
the complete NLO contribution. We note that the
invariant $\tr( \phi \phi G)$ connects the field insertion $\sim \phi_a
\phi_b$ with a single propagator line. If a diagram contains that
invariant more than once it is always possible to disconnect the graph by
cutting two lines.\footnote{
To see this, note that the number of loops $L$ in these diagrams is given
by the standard relation
\beq
 L = I - V_3 - V_4 + 1 = V_4 + \frac{V_3}{2} + 1,
\eeq
where we have used Eq.\ (\ref{lines}) for the second equality. Tadpole
diagrams are forbidden in those graphs and each ``bubble'' $\tr( G^2 )$
corresponds to one closed loop.  From the above relation one then observes
that the total number of loops in the diagram is given by the number of
bubbles plus one. In particular, one can disconnect the diagram by cutting
two lines connecting the fields insertions (the terms $\sim \tr( \phi \phi
G)$).} Consequently, a diagram with the above structure and with $V_3
\geqslant 4$ is not two-particle irreducible and does not contribute to
$\Gamma[\phi,G]$. We therefore have classified all possible diagrams which
contribute to the $2PI$ effective action at NLO and present the result in
the next section. In Appendix A we discuss possible further approximations
consistent with an expansion in powers of $1/N$.

\section{The $2PI$ effective action at NLO}
\label{2PINLOaction}

We write 
\beq
\Gamma_2[\phi,G]= \Gamma_2^{\rm LO}[G] 
          + \Gamma_2^{\rm NLO}[\phi,G] 
          + \ldots
\eeq
where $\Gamma_2^{\rm LO}$ denotes the leading order (LO) 
and $\Gamma_2^{\rm NLO}$ the next-to-leading order (NLO) contributions. 
The LO contribution to $\Gamma_2[G]$ is given by the diagram presented in 
Fig.\ \ref{LOfig}, 
\beql{LOcont}
\Gamma_2^{\rm LO}[G] = -\frac{\lambda}{4!N} \int_{x} G_{aa}(x,x)G_{bb}(x,x).
\eeql   
This contribution is $\phi$--independent. 

The NLO contribution consists of an infinite series of diagrams which fall
into two classes. The first class is independent of $\phi$ and is
constructed with only quartic vertices. It contains the complete NLO
contribution when $\phi = 0$ and has been resummed in Ref.\ 
\cite{Berges:2001fi}. The diagrams are shown in Fig.~\ref{NLOfig}.  
The second class depends on $\phi$ and contributes for the case of a 
nonzero field expectation value. This series of diagrams is shown in Fig.\
\ref{NLOfigb} and can be resummed as well. As a result, the NLO
contribution to $\Gamma_2[\phi,G]$ is given by the sum of the two resummed
expressions and reads\footnote{
Besides the dynamical field degrees of
freedom $\phi$ and $G$ we will introduce a number of quantities which are
(resummed) functions of these fields. These functions will be denoted by
either boldface or Greek letters in the following.
}
\beq
\label{NLOcont} 
\Gamma_2^{\rm NLO}[\phi,G] =  \frac{i}{2} \Tr_{_\C}  
\mbox{Ln} [\, {\bf B}(G)\, ]  \nonumber \\ 
+\frac{i\lambda}{6N} \int_{xy} 
{\bf I}(x,y;G) \phi_a(x) G_{ab}(x,y) \phi_b (y). 
\eeq
In the above equation we have defined
\beq
\label{Feq}
{\bf B}(x,y;G) = \delta_{\C}(x-y)
 + i \frac{\lambda}{6 N} G_{ab}(x,y)G_{ab}(x,y),
\eeq
and the logarithm in Eq.\ (\ref{NLOcont}) sums the infinite series shown 
in Fig. \ref{NLOfig}, 
\bea
&&
\Tr_{_\C} \mbox{Ln}[\, {\bf B}(G)\, ] 
=  \int_{x} 
 \left( i \frac{\lambda}{6 N} G_{ab}(x,x)G_{ab}(x,x) \right)
\nonumber\\
&& \quad -\frac{1}{2} \int_{xy}
\left( i \frac{\lambda}{6 N} G_{ab}(x,y)G_{ab}(x,y) \right)
\left( i \frac{\lambda}{6 N}\, G_{a'b'}(y,x)G_{a'b'}(y,x) \right)
\nonumber\\
&& \quad + \ldots
\eea
The function ${\bf I}(x,y;G)$ is defined by 
\beq
\label{ieqab}
{\bf I} (x,y;G) = \frac{\lambda}{6 N} G_{ab}(x,y) G_{ab}(x,y)
 - i \frac{\lambda}{6 N} \int_{z} {\bf I}(x,z;G)
 G_{ab}(z,y) G_{ab}(z,y) \, ,
\eeq
and resums the infinite chain of ``bubble'' graphs, which can be seen by 
re-expanding the series. The functions ${\bf I} (x,y;G)$ and the inverse 
of ${\bf B}(x,y;G)$ are closely related by
\beql{Binverse}
{\bf B}^{-1}(x,y;G) = \delta_{\C}(x-y) - i {\bf I} (x,y;G) \, ,
\eeql
which follows from convoluting Eq.\ (\ref{Feq}) with ${\bf B}^{-1}$ and
using Eq.\ (\ref{ieqab}). We note that $\bf B$ and $\bf I$ do not depend
on $\phi$, and $\Gamma_2[\phi,G]$ is only quadratic in $\phi$ at NLO.
Hence, the complete effective action at NLO contains only quadratic and
quartic terms in $\phi$.

\section{The equations of motion}
\label{Secteom}

From the $2PI$ effective action $\Gamma[\phi,G]$ at NLO we find 
equations of motion for the macroscopic fields $\phi$ and $G$, as 
indicated in Sec. \ref{section2pi}. The equation for the field expectation 
value (\ref{feeq}) reads at NLO
\beq
\label{directeqphi}
 -\left(\square_x + m^2 + \frac{\lambda}{6N}\left[ \phi^2(x) + 
 G_{cc}(x,x) \right]\right) \phi_a(x)  =  {\bf K}_a(x,x).
\eeq
We have written the LO contribution of the evolution equation
on the LHS and combined the NLO contribution as\footnote{Note that the 
classical inverse propagator $iG_0^{-1}$ contains a LO and a NLO 
part.}
\beql{KK}
 {\bf K}_a(x,y) \equiv
 {\bf K}_a(x,y;\phi,G) \equiv  \frac{\lambda}{3N} \int_z \, 
 {\bf B}^{-1}(x,z;G) G_{ab}(y,z)\phi_b(z), 
\eeql
evaluated at $x=y$. We have written ${\bf K}_a(x,y)$ as a 
function of $x$ and $y$ for later convenience. 

Eq.\ (\ref{gap}) for $G^{-1}_{ab}(x,y)$ can be completed using Eq.\ 
(\ref{eqself}) for the self-energy. The LO contribution is simply 
\begin{equation}
\Sigma^{\rm LO}_{ab}(x,y) = -i \frac{\lambda}{6N} G_{cc}(x,x) 
\delta_{ab}\delta_\C(x-y). 
\end{equation}
To obtain the NLO contribution the following identity may be helpful
\begin{equation}
\frac{\delta {\bf I}(u,v;G)}{\delta G_{ab}(x,y)} = 
i\frac{\delta {\bf B}^{-1}(u,v;G)}{\delta G_{ab}(x,y)} = 
\frac{\lambda}{3N} {\bf B}^{-1}(u,x;G) G_{ab}(x,y) {\bf B}^{-1}(y,v;G),
\end{equation}
where we used that
\begin{equation}
\frac{\delta {\bf B}(u,v;G)}{\delta G_{ab}(x,y)} =
i\frac{\lambda}{3N} G_{ab}(x,y) \, \delta_\C(u-x)\delta_\C(v-y).
\end{equation}
Collecting all the pieces, we find that Eq.\ ({\ref{gap}) for the inverse 
propagator can be written as
\bea
\label{directeqG}
iG_{ab}^{-1}(x,y) =   
-\left(\square_x + m^2 + \frac{\lambda}{6N} \left[\phi^2(x) + 
  G_{cc}(x,x)\right] \right) \delta_{ab} \delta_{\C}(x-y) &&
 \nonumber\\ 
 - \frac{\lambda}{3N} {\bf B}^{-1}(x,y;G) \phi_a(x)\phi_b(y)
 + i{\bf D} (x,y) G_{ab}(x,y), &&
\label{gevol}
\eea
with the definition
\beal{DD}
\nonumber
&& \hspace{-0.5cm} {\bf D}(x,y) \equiv {\bf D}(x,y;\phi,G) \equiv 
i\frac{\lambda}{3N}{\bf B}^{-1}(x,y;G)  \\
 && \quad +  \left(\frac{\lambda}{3N}\right)^2 
 \int_{uv} {\bf B}^{-1}(x,u;G) 
 \phi_a(u)G_{ab}(u,v)\phi_b(v){\bf B}^{-1}(v,y;G).
\eeal
Since ${\bf B}^{-1}$ is of order one and ${\bf D}$ of order $1/N$, 
the first line on the RHS of Eq.\ (\ref{gevol}) corresponds to
the LO and the second line to the NLO contribution.

Eqs.\ (\ref{directeqphi}, \ref{gevol}) together with Eqs.\ (\ref{KK}, 
\ref{DD}) and (\ref{ieqab}, \ref{Binverse})
form the complete set of equations which have to be solved to 
obtain the 
$2PI$ effective action at NLO in the $1/N$ expansion. From 
$\Gamma[\phi,G]$ all correlation functions can then be found by 
derivatives with respect to the fields as functions of the known $\phi$ 
and $G$. We stress that the $2PI$--$1/N$ expansion is done on the level 
of the effective action. There are no further approximations involved on
the level of the evolution equations. 

We note that Eq.\ (\ref{DD}) contains a double integration over the
time contour $\C$ which can be inconvenient for numerical purposes.
It turns out that it is possible to disentangle the nested integrations by 
exploiting the function ${\bf K}_a$. Convoluting the functions ${\bf B}$ 
and ${\bf D}$ and using the definitions for ${\bf B}$
and ${\bf K}_a$ in Eqs.\ (\ref{Feq}) and (\ref{KK}), 
one obtains 
\beal{DDnew}
 {\bf D} (x,y)
&=&  i \frac{\lambda}{3N} \delta_{\C}(x-y) 
+ \frac{\lambda}{3N} {\bf K}_a (y,x) \phi_a(x) \nonumber \\&&
 -i \frac{\lambda}{6N} \int_z G_{ab}(x,z) G_{ab}(x,z) {\bf D}(z,y). 
\eeal 
We observe that the nested integrals   
have disappeared in Eq.\ (\ref{DDnew}) without any problems. In Appendix B we 
work out more details for the equations preserving the nested-integral 
structure. It is also convenient to rewrite Eq.\ (\ref{KK}) for ${\bf 
K}_a$ such that ${\bf B}$ does not appear. By convoluting ${\bf B}$ and 
${\bf K}_a$, one obtains  
\begin{equation}
\label{Knew}
 {\bf K}_a (x,y) = \frac{\lambda}{3N} \phi_b(x) G_{ba}(x,y) -
 i \frac{\lambda}{6N} \int_z G_{bc}(x,z) G_{bc}(x,z) {\bf K}_a(z,y). 
\end{equation}
As a result, we find that ${\bf B}$ and ${\bf B}^{-1}$ are eliminated 
completely from the coupled equations. 
We also see that the gap equations for ${\bf D}$ and ${\bf K}_a$ 
are local in one time variable (here $y$), which is useful for numerical 
implementation.

The form of the equation of motion for the inverse propagator, Eq.\
(\ref{gevol}), is suitable for a boundary value problem and can be used,
in particular, to discuss the propagator in thermal equilibrium by
specifying the contour $\C$ to the Matsubara contour along the
imaginary-time axis. However, to deal with nonequilibrium time evolution,
i.e., an initial value problem, the form already given in Eq.\
(\ref{geeq}) is more useful. At NLO we find that the partial differential
equation for the propagator takes the form
\beal{Gnew}
 -\left( \square_x+m^2+\frac{\lambda}{6N}
\left[ \phi^2(x) +  G_{cc}(x,x) \right] \right) G_{ab}(x,y) =
 i\delta_{ab} \delta_\C(x-y)
\nonumber && \\
+  \phi_a (x) {\bf K}_b(x,y) - 
 i \int_z {\bf D}(x,z) G_{ac}(x,z) G_{cb}(z,y). &&
\eeal
To summarize, Eqs.\ (\ref{directeqphi}, \ref{Gnew}) together with Eqs.\ 
(\ref{DDnew}, \ref{Knew}) form the complete set of equations which have to 
be solved. They are completely equivalent to our first set containing 
Eq.\ (\ref{gevol}) for the inverse propagator.

\section{The auxiliary-field method}
\label{auxiliary}

In this section we show that the equations derived in the previous
sections can also be obtained in the auxiliary-field formulation. 
In particular, we
discuss the difference between the $2PI$--$1/N$ expansion and the
``bare-vertex approximation'' introduced in Ref.\ \cite{Mihaila:2000sr}.

Following Refs.\ \cite{Mihaila:2000sr,Coleman:jh} we 
rewrite the action by introducing an auxiliary field $\chi$ as
\beql{auxfieldaction}
S[\varphi,\chi] = -\int_x \left[ \half \varphi_a(\square+m^2)\varphi_a
 -\frac{3N}{2\lambda}\chi^2 + \frac{1}{2}\chi\varphi_a\varphi_a  \right].
\eeql
Integrating out $\chi$ yields the original action (\ref{classical}).
From the Heisenberg equations of motion we see that the 
auxiliary field represents the composite operator $\chi(x) = \lambda/(6N) 
\varphi_a(x)\varphi_a(x)$.
The following one- and two-point 
functions can be written down:
\beq
 \phi_a(x) = \bra \varphi_a(x)\ket,\;\;\;\;\;\;\;\;
 \bar\chi(x) =  \bra \chi(x)\ket, \\
\eeq
and
\bea
\nonumber
G_{ab}(x,y) &=& \bra T_{\cal C} \varphi_a(x)\varphi_b(y) \ket -
 \bra \varphi_a(x)\ket\bra \varphi_b(y)\ket,\\
{\bf K}_{a}(x,y) &=&  \bra T_{\cal C} \chi(x)\varphi_a(y) \ket -
 \bra \chi(x)\ket\bra \varphi_a(y)\ket = \bar {\bf K}_a(y,x),\\
{\bf D}(x,y) &=& \bra T_{\cal C} \chi(x)\chi(y) \ket -
 \bra \chi(x)\ket\bra \chi(y)\ket.
\nonumber
\eea
We note that additional ``propagators'' appear due to the introduction of 
the auxiliary field. Since $\chi$ is not a dynamical degree of 
freedom, only $G_{ab}$ has a physical meaning of a propagator. The role 
of the other one- and two-point functions is to implicitly perform 
infinite resummations, which were carried out explicitly in the direct 
method in the previous sections.

Following \cite{Mihaila:2000sr} the quantum fields are combined in an 
extended field with $N+1$ components, 
\beq 
\Phi_i =  \left( \begin{array}{c} \varphi_a \\ \chi \end{array} \right),
\eeq
where $i=1,\ldots, N+1$. 
One may now formulate the $2PI$ effective action for this quantum field
theory by coupling sources to the field $\Phi_i(x)$ and the bilocal field
$\Phi_i(x)\Phi_j(y)$ \cite{Mihaila:2000sr}. We would like to point out
that in the presence of sources the original quantum theory and the one
with the auxiliary field are potentially different. For instance, in the
second formalism one may differentiate with respect to the bilocal source and
obtain correlation functions of $\chi(x)\varphi_a(y)$. This possibility
is not present in the original theory. However, as we will see below, on
the level of the equations of motion in the $2PI$--$1/N$ expansion at
NLO, the two approaches yield identical results.

The inverse of the classical propagator is
\begin{equation}
i \mathcal{G}^{-1}_{0,ij}(x,y;\bar\Phi) = 
 \frac{\delta^2 S[\bar\Phi]}{\delta\bar\Phi_i(x) \delta\bar\Phi_j(y)},
\end{equation}
where $\bar\Phi_i = (\phi_a, \bar{\chi})$ and $S[\bar\Phi]=S[\phi,\bar\chi]$. 
It has the following components:
\bea
\nonumber
 \frac{\delta^2 S[\phi,\bar\chi]}{\delta\phi_a(x)\delta\phi_b(y)} &=& 
 -\left[\square_x+m^2+\bar\chi(x)\right]\delta_{ab}\delta_{\C} (x-y),\\
 \frac{\delta^2 S[\phi,\bar\chi]}{\delta\bar\chi(x)\delta\bar\chi(y)} &=&
 \frac{3N}{\lambda}\delta_{\C} (x-y),\\
 \frac{\delta^2 S[\phi,\bar\chi]}{\delta\bar\chi(x)\delta\phi_a(y)} &=&
 - \phi_a(x)\delta_{\C} (x-y).
\nonumber
\eea
These operators are symmetric. 
Similarly, the matrix containing the two-point functions is defined as
\beq
 \mathcal{G}_{ij} = \left(
 \begin{array}{cc} G_{ab} & \bar {\bf K}_a \\
 {\bf K}_b & {\bf D} \end{array} \right),
\eeq
where $\bar{\bf K}_a(x,y)={\bf K}_a(y,x)$ \cite{Mihaila:2000sr}. Hence, 
this matrix is also symmetric.

The $2PI$ effective action can now be written down and reads
\beq
 \Gamma [\bar\Phi,\mathcal{G}] = 
 S[\bar\Phi] + \frac{i}{2}\Tr\,\ln \mathcal{G}^{-1} + 
 \frac{i}{2}\Tr\,\mathcal{G}^{-1}_0\mathcal{G} + \Gamma_2[\mathcal{G}]
 +\mbox{const.}
\eeq
Here $\Gamma_2$ is given by all two-particle irreducible graphs made with
lines representing the ``propagators'' $G_{ab}$, ${\bf K}_a$, $\bar{\bf 
K}_a$, and ${\bf D}$,
and the vertex $-\half \chi(x)\varphi_a(x)\varphi_a(x)$. In the
auxiliary-field formalism, $\Gamma_2$ does not depend on the expectation 
value $\bar\Phi_i$.

The equations of motion for the field expectation values follow by
variation of $\Gamma$ with respect to $\phi_a$ and $\bar\chi$. We find
\beql{auxfieldeq1}
 -\left[ \square_x+m^2+\bar\chi(x)\right] \phi_a(x) = 
 \frac{1}{2}\left[ {\bf K}_a(x,x) + \bar  {\bf K}_a(x,x)\right]
 = {\bf K}_a(x,x), 
\eeql
and
\beql{auxfieldeq2}
 \bar\chi(x) = \frac{\lambda}{6N}\left[\phi^2(x) + G_{cc}(x,x) 
 \right].
\eeql
The equation for the two-point function follows by variation with
respect to $\mathcal{G}$, which gives
\beq
 \mathcal{G}^{-1}_{ij} = \mathcal{G}^{-1}_{0,ij} - 
 2i\frac{\delta \Gamma_2[\mathcal{G}]}{\delta \mathcal{G}_{ij}}.
\eeq
By convoluting this equation from the right with $\mathcal{G}$
and decomposing the self-energy as \footnote{Note that in the 
auxiliary-field formalism $\hat\Sigma_{ab}$ does not receive a local LO  
contribution. It differs therefore from the self energy $\Sigma_{ab}$ in 
the direct method, obtained by varying $\Gamma_2[\phi,G]$.}
\beq
 2i\frac{\delta \Gamma_2[\mathcal{G}]}{\delta \mathcal{G}_{ij}} = 
 \left(
 \begin{array}{cc} \hat\Sigma_{ab} & \bar\Xi_a \\
 \Xi_b & \Pi \end{array} \right),
\eeq
one obtains the following set of coupled equations:
\bea
 &&-\left[ \square_x + m^2 +\bar\chi(x) \right] G_{ab} (x,y) =
 \phi_a(x) {\bf K}_{b}(x,y) + i\delta_{ab} \delta_\C (x-y) 
\nonumber
\\
 &&\hspace{1.5cm} +
 i \int_z \left[ \hat\Sigma_{ac} (x,z) G_{cb} (z,y) + 
 \bar\Xi_{a} (x,z) {\bf K}_{b} (z,y)\right],
 \label{eqGab}
\\
 &&\frac{3N}{\lambda} {\bf K}_a(x,y) = \phi_b(x) G_{ba} (x,y) 
 \nonumber
\\
 &&\hspace{1.5cm} +
 i\int_z \left[ \Xi_{b} (x,z) G_{ba} (z,y) + 
 \Pi(x,z) { \bf K}_{a} (z,y) \right],
\label{eqK}
\\
 && \frac{3N}{\lambda} {\bf D}(x,y) = 
 \phi_a(x) \bar {\bf K}_a(x,y) + i\delta_\C(x-y)
 \nonumber\\
 &&\hspace{1.5cm} +
 i\int_z\left[\Xi_{a}(x,z)\bar {\bf K}_{a}(z,y) + 
 \Pi (x,z) {\bf D}(z,y) \right],
\label{eqD2}
\\
 &&-\left[ \square_x + m^2 + \bar\chi(x) \right] \bar {\bf K}_{a} (x,y) =
 \phi_a(x) {\bf D}(x,y) 
\nonumber\\
 &&\hspace{1.5cm} +
 i \int_z\left[\hat\Sigma_{ab}(x,z)\bar {\bf K}_{b} (z,y) + 
 \bar\Xi_{a} (x,z) {\bf D}(z,y) \right].
 \label{eqbarK2} 
\eea
We note that Eq.\ (\ref{eqbarK2}) for $\bar {\bf K}_{a}$ is not an
independent equation since $\bar {\bf K}_a(x,y) = {\bf K}_a(y,x)$.
Therefore, Eq.\ (\ref{eqbarK2}) is not needed in practice.

\begin{figure}[t]
\centerline{\psfig{figure=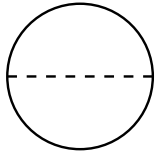,height=2cm}}
\caption{
NLO contribution in the $2PI$--$1/N$ expansion in the auxiliary-field
formalism.  The full line denotes the scalar propagator $G_{ab}$, the
dashed line the auxiliary-field propagator ${\bf D}$.
}
\label{figauxNLO}
\end{figure}

To find explicitly at which order in the $2PI$--$1/N$ expansion specific
diagrams contribute, we note that in the auxiliary-field formalism the 
possible irreducible $O(N)$--singlets are of the form
\begin{equation}
\tr(G^n), \;\;\;\; {\bf D}, \;\;\;\; \tr({\bf K}G^n\bar{\bf K}),
\end{equation}
with $n\geq 1$. From Eq.\ (\ref{eqD2}) it follows that ${\bf D} \sim 1/N$, 
and from  Eq.\ (\ref{eqK}) we find that
\begin{equation}
 \tr ({\bf K}G^n \bar{\bf K}) \sim \frac{1}{N^2}\tr( \phi\phi G^{n+2})
\eeq
is proportional to $1/N$ as well. Using this scaling behaviour it is 
straightforward to give the diagrams that contribute at NLO and NNLO in 
the $2PI$--$1/N$ expansion in the auxiliary-field formalism.

We find that the NLO contribution to $\Gamma_2$ consists of one graph 
only, 
\beq
 \Gamma_2^{\rm NLO} [\mathcal{G}] = \frac{i}{4} \int_{xy} G_{ab}(x,y) 
 G_{ab}(x,y) {\bf D}(x,y),
\label{NLOA}
\eeq
shown in Fig.\ \ref{figauxNLO}.
{}From this expression the self-energies defined above follow:
\bea
\nonumber
 \hat\Sigma_{ab}^{\rm NLO}(x,y) &=& -G_{ab}(x,y){\bf D}(x,y),\\
 \Pi^{\rm NLO}(x,y) &=& -\frac{1}{2}G_{ab}(x,y)G_{ab}(x,y), \\
 \Xi_a^{\rm NLO}(x,y) &=& 0.
\nonumber
\eea
Inserting these expressions in Eqs.\ (\ref{eqGab})--(\ref{eqD2}), we 
immediately recover our final result at NLO, Eqs.\ (\ref{directeqphi}, 
\ref{DDnew},  \ref{Knew}, \ref{Gnew}), obtained by the direct method in 
Sec.~\ref{Secteom}.\footnote{Therefore, we use the same notation 
for the functions ${\bf K}$, ${\bf D}$ and $\Pi$ in Sects.\
\ref{Secteom} to \ref{secevolution}.} 
\begin{figure} 
\centerline{\psfig{figure=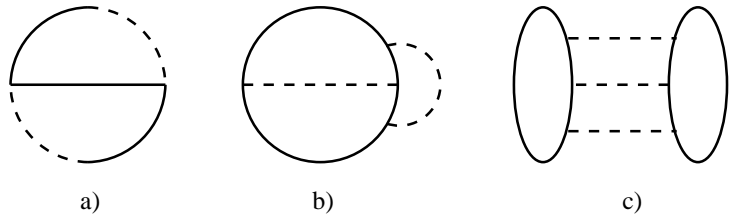,height=2.8cm}}
\caption{ 
NNLO contribution in the $2PI$--$1/N$ expansion in the auxiliary-field 
formalism. The full-dashed lines denote the mixed propagators ${\bf K}_a,
\bar{\bf K}_a$.
}
\label{nextfig2pi}
\end{figure}

Only three diagrams contribute in the auxiliary-field formulation at NNLO.  
They are shown in Fig.\ \ref{nextfig2pi}. We note that diagrams with the
mixed propagator ${\bf K}_a$ (resulting in a nonvanishing $\Xi_a$ and
$\bar\Xi_a$) appear only at NNLO.
In Ref.\ \cite{Mihaila:2000sr} the first NNLO diagram, 
\begin{equation}
 \Gamma_2^{\rm NNLO a } [\mathcal{G}] =
\frac{i}{2} \int_{xy} {\bf K}_{a}(x,y) G_{ab}(x,y) \bar{\bf K}_b(x,y),
\end{equation}
was combined with the NLO diagram of Fig.\ \ref{figauxNLO} in the 
so-called bare-vertex approximation (BVA). We conclude that in the 
presence of a field expectation value the BVA approximation is not 
consistent with the $2PI$--$1/N$ approximation discussed 
here. For vanishing $\phi$, the $2PI$--$1/N$ approximation at NLO and 
the BVA ansatz are identical.\footnote{For similar approximation
schemes see also Ref.\ \cite{Lutz}.}

\section{Evolution equations for the spectral and statistical functions
at NLO}
\label{secevolution}

In order to describe nonequilibrium dynamics we will now specify the 
contour $\C$ to the standard Schwinger-Keldysh contour along the real-time 
axis \cite{Schwinger:1961qe}.\footnote{We use a Gaussian initial 
density matrix. Non-gaussian initial density matrices are also 
possible, see e.g.\ Refs.\ \cite{Calzetta:1986cq,Berges:2000ur}.}
The two-point function can be decomposed as
\bea
 G_{ab}(x,y)&=&G^>_{ab}(x,y) \Theta_{\C}(x^0-y^0)+
 G^<_{ab}(x,y) \Theta_{\C}(y^0-x^0),
\eea
where $G^{>}_{ab}(x,y) = {G^{<}_{ab}}^*(x,y)$ are complex functions. 
For the real scalar field theory it is convenient to express 
the evolution equations in terms of two independent real--valued 
two--point functions, which can be associated to the expectation values of 
the commutator and the anti-commutator of two fields 
\cite{Aarts:2001qa,Berges:2001fi,Aarts:2001yn}. We define  
\bea
F_{ab} (x,y) &=& \frac{1}{2} \Big(G^>_{ab}(x,y)+G^<_{ab} (x,y)\Big) 
= {\rm Re}[G^>_{ab} (x,y)]
\label{defrho},\\
\rho_{ab} (x,y) &=& i \Big(G^>_{ab}(x,y)-G^<_{ab} (x,y)\Big) 
 = -2 {\rm Im}[G^>_{ab} (x,y)].
\eea
Here $F$ is the statistical propagator and $\rho$ denotes the 
spectral function, with the properties $F^*_{ab}(x,y)=F_{ab}(x,y)=F_{ba}(y,x)$ 
and $\rho^*_{ab}(x,y)=\rho_{ab}(x,y)=-\rho_{ba}(y,x)$. 

In order to proceed it is convenient to separate the singular part of
${\bf D}$ (see Eq.\ (\ref{DDnew})) and write
\begin{equation}
{\bf D}(x,y) = \frac{\lambda}{3N}\left[ 
i\delta_{\C}(x-y)+ {\bf {\hat D}}(x,y) \right],
\end{equation}
with
\begin{equation}
\label{eqDhat}
 {\bf {\hat D}} (x,y) = {\bf K}_a (y,x) \phi_a(x)
 -\frac{\lambda}{3N} \Pi(x,y)
 + \frac{i\lambda}{3N} \int_z  \Pi(x,z)
{\bf {\hat D}} (z,y).
\end{equation}
For the functions ${\bf K}_a(x,y)$ (see Eq.\ (\ref{Knew})) and 
${\bf\hat D}(x,y)$ we define the statistical and spectral components as
\bea
 {\bf K}^F_{a} (x,y) & = &  
 \frac{1}{2} \Big( {\bf K}_a^>(x,y)+ {\bf K}_a^<(x,y) \Big) = 
 {\rm Re}[\,{\bf K}_a^> (x,y)] \label{defF},\\
 {\bf K}_{a}^{\rho} (x,y) & = &
 i \Big( {\bf K}_a^> (x,y)- {\bf K}_a^< (x,y) \Big) = 
 -2\, {\rm Im}[\,{\bf K}_a^>(x,y)],
\eea
and the same for ${\bf\hat D}_F(x,y)$ and ${\bf\hat D}_\rho(x,y)$.

Now we have all the necessary definitions and relations to express the
time evolution equations for the field expectation value and Green's
function along the Schwinger-Keldysh contour as real and causal equations.  
The time evolution equation for the field reads
\begin{equation}
-\left[\square_x + m^2 + \bar\chi(x)\right] \phi_a(x)  =  {\bf K}^F_{a}(x,x)
\label{Rphi}
\end{equation}
with
\begin{equation}
 \bar\chi(x) =  \frac{\lambda}{6N}\left[ \phi^2(x) + F_{cc}(x,x) \right]. 
\end{equation}
The functions ${\bf K}_{a}^F$ and ${\bf K}_{a}^{\rho}$ satisfy the 
equations 
\bea
{\bf K}_{a}^F(x,y) = \frac{\lambda}{3N}\phi_b(x) F_{ba}(x,y) + 
\frac{\lambda}{3N}\int_{0}^{x^0} dz\, \Pi_\rho(x,z){\bf K}_a^F(z,y) 
&& \nonumber\\
- \frac{\lambda}{3N} \int_0^{y^0} dz\, \Pi_F(x,z) {\bf K}_a^\rho(z,y),
&&  \label{IRFR}
\\
{\bf K}_{a}^{\rho}(x,y) = \frac{\lambda}{3N}\phi_{b}(x)\rho_{ba}(x,y)
+ \frac{\lambda}{3N} \int_{y^0}^{x^0} dz\, \Pi_\rho(x,z){\bf 
K}_a^\rho(z,y), &&  \nonumber
\eea
where we employ the notation
\beql{k2}
 \int^{x^0}_0 dz \equiv \int^{x^0}_0 dz^0 \int d{\bf z} \, ,
\eeql
and
\bea
 \Pi_F (x,y) & = & - \frac{1}{2}\left[ F_{ab}(x,y) F_{ab}(x,y) -
 \frac{1}{4}\rho_{ab}(x,y) \rho_{ab}(x,y) \right],
 \nonumber\\
 \Pi_{\rho} (x,y) & = & - F_{ab} (x,y) \rho_{ab} (x,y).
\eea
The statistical propagator obeys
\bea 
 \nonumber
&& -\left[\square_x + m^2 +\bar\chi(x)\right] F_{ab}(x,y)  = 
 \frac{\lambda}{3N} F_{ac}(x,x) F_{cb}(x,y) 
 + \phi_a(x) {\bf K}_{b}^F(x,y) \\
&& \quad\quad\quad\quad
 +\int_0^{x^0} dz \, \hat\Sigma_{ac}^\rho(x,z) F_{cb}(z,y) 
 -\int_0^{y^0} dz \, \hat\Sigma_{ac}^F(x,z)\rho_{cb} (z,y),
 \label{KGreen} 
\eea
and the spectral function
\bea 
 \nonumber 
-\left[\square_x + m^2 + \bar\chi(x) \right] \rho_{ab}(x,y)  &=& 
 \frac{\lambda}{3N} F_{ac}(x,x) \rho_{cb}(x,y) 
+ \phi_a(x) {\bf K}_{b}^{\rho} (x,y) \\
&& + \int_{y^0}^{x^0} dz \, \hat\Sigma^\rho_{ac}(x,z)\rho_{cb}(z,y). 
 \label{Kspectral}
\eea
Here we use the notation
\bea
&& \hat\Sigma_{ab}^F(x,y) = -\frac{\lambda}{3N}\left[ 
 F_{ab}(x,y)\hat {\bf D}_F(x,y) - \frac{1}{4} \rho_{ab}(x,y)\hat {\bf 
 D}_\rho(x,y)\right],\\
&& \hat\Sigma_{ab}^\rho(x,y) = -\frac{\lambda}{3N}\left[ 
 \rho_{ab}(x,y)\hat {\bf D}_F(x,y) + F_{ab}(x,y)\hat {\bf 
 D}_\rho(x,y)\right],
\eea
with
\bea
 {\bf {\hat D}}_{ F}(x,y) &=&  
 {\bf K}_{a}^F(y,x)  \phi_a (x)  -
 \frac{\lambda}{3N}\Pi_F (x,y) \nonumber\\ 
 && \hspace{-0.5cm} +  \frac{\lambda}{3N} \, \int_{0}^{x^0} dz\,
 {\Pi}_{\rho}(x,z) \, {\bf {\hat D}}_{ F} (z,y)
 -  \frac{\lambda}{3N} \, \int_{0}^{y^0} dz\,
 {\Pi}_{F}(x,z) \, {\bf {\hat D}}_{\rho} (z,y), 
 \nonumber\\
 {\bf {\hat D}}_{\rho}(x,y) &=& 
 -\, {\bf K}_{a}^{\rho}(y,x)  \phi_a (x) -
 \frac{\lambda}{3 N} \Pi_{\rho} (x,y)
 +  \frac{\lambda}{3N} \, \int_{y^0}^{x^0} dz\, 
 {\Pi}_{\rho}(x,z) \, {\bf {\hat D}}_{\rho} (z,y). 
 \nonumber\\ 
 \label{DRFR}
\eea
In absence of a field-expectation value ($\phi_a=0$) we find that 
${\bf K}_a^F = {\bf K}_a^\rho = 0$ and the equations above reduce to those
treated in \cite{Berges:2001fi,Aarts:2001yn} for diagonal two-point
functions.

In order to completely determine the time evolution, 
Eqs.\ (\ref{Rphi}), (\ref{KGreen}) and (\ref{Kspectral})
have to be implemented with initial conditions taken at $x^0 = y^0 = 0$.
For the field $\phi_a(x)$ one may choose nonvanishing
values $\phi_a(x^0 = 0,{\bf x})$, but vanishing "velocities"
$\partial_{x^0} \phi_a(x)|_{x^0=0} = 0$. The initial values for the 
spectral function are completely fixed by the equal-time 
properties \cite{Aarts:2001qa} 
\bea
\rho_{ab}(x,y) \vert_{x^0 = y^0} = 0, \;\;\;\;
\partial_{x^0} \rho_{ab} (x,y)\vert_{x^0 = y^0}
 = \delta_{ab} \delta^d({\bf x-y}).
\eea
Nontrivial information about the initial density matrix is contained in 
(derivatives of) the statistical two-point function at initial time
\begin{equation}
F_{ab}(x,y)\vert_{x^0 = y^0=0}, \;\;\;\; 
\partial_{x^0} F_{ab}(x,y)\vert_{x^0 = y^0=0},  \;\;\;\;
\partial_{x^0} \partial_{y^0} F_{ab}(x,y)\vert_{x^0 = y^0=0}.
\end{equation}
Specification of these three functions is necessary and sufficient to 
solve the equations of motion.

\section{Weak coupling expansion }
\label{weakcoupling}

In order to simplify the interrelated set of nonperturbative NLO equations
of motion given in the previous Section, we discuss here a truncated
version of the $2PI$--$1/N$ approximation for the case that the coupling
is weak. It amounts to expanding the effective action
to second order in explicit factors of the coupling constant $\lambda$.
Since $\Gamma_2^{\rm LO}$ is proportional to $\lambda$ it is preserved 
completely. 
For the NLO contribution $\Gamma_2^{\rm NLO}$ in Eq.~(\ref{NLOcont}) we 
find
\bea
\Gamma_2^{\rm NLO}[\phi,G] &\simeq & \frac{\lambda}{6N} \int_x \Pi(x,x)
   + i\left(\frac{\lambda}{6N}\right)^2 \int_{xy} \Pi(x,y) \Pi(x,y)
\nonumber \\
&&  -2i\left(\frac{\lambda}{6N}\right)^2 \int_{xy} \Pi(x,y)
\phi_a(x) G_{ab}(x,y) \phi_b(y),
\label{twoloop}
\eea
where we use again the notation $\Pi(x,y) = 
-\frac{1}{2}G_{ab}(x,y)G_{ab}(x,y)$.
The cor\-respon\-ding diagrams are presented in Fig.\ 
\ref{figweak}.\footnote{Here we consider the simple case of a field 
expectation value for which $(\lambda/6N) \phi^2$ is small compared to
the characteristic mass scale. We stress that for a consistent weak
coupling expansion it is important to note that a nonzero minimum at
$\phi^2 =\phi_0^2$ of the classical potential in (\ref{classical}) scales
as $\phi_0^2 = -(6N/\lambda) m^2$. Therefore, in a situation with
spontaneously broken symmetry it is not sufficient to count only the
powers of $\lambda$ coming from the vertices, as exemplified in this
section. A consistent ${\cal O} (\lambda)$ scheme would have to take into
account the first and the third graph of Fig.\ \ref{figweak}, while at
${\cal O}(\lambda^2)$ the three-loop graphs of Figs.\ \ref{figweak} and
\ref{NLOfigb} would have to be taken into account.}

\begin{figure} 
\centerline{
\psfig{figure=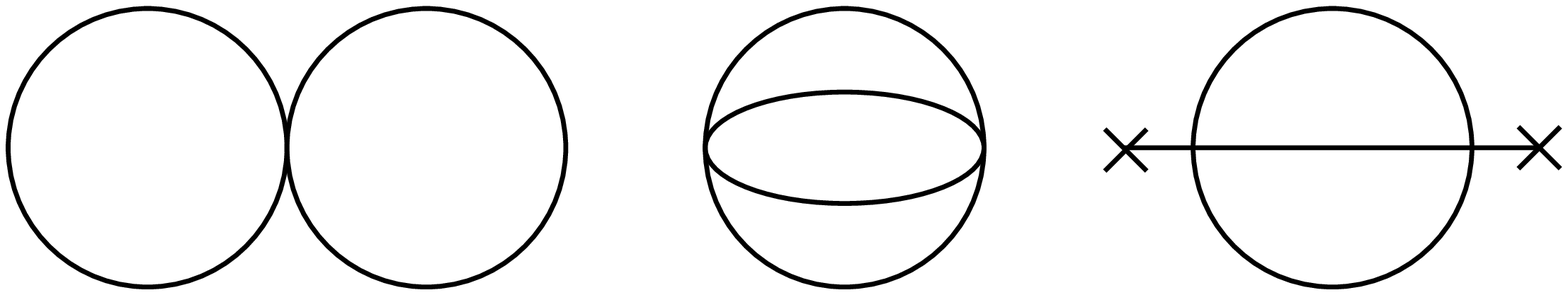,width=9cm} 
}
\caption{ 
Diagrams contributing in the $2PI$--$1/N$ expansion when an additional 
weak-coupling expansion to second order in $\lambda$ is performed.
}
\label{figweak}
\end{figure}

The weak-coupling expansion affects the equations of motion through  
the auxiliary variables ${\bf K}_a$ and ${\bf \hat 
D}$, 
\beal{eqweak}
{\bf K}_a(x,y) & \simeq & \frac{\lambda}{3N}\phi_b(x)G_{ba}(x,y) 
+ i \left( \frac{\lambda}{3N} \right)^2\int_z 
\Pi(x,z)\phi_b(z)G_{ba}(z,y),
\nonumber\\
{\bf \hat D}(x,y) & \simeq & \frac{\lambda}{3N} 
\left[ \phi_a(x)G_{ab}(x,y)\phi_b(y) - \Pi(x,y)\right].
\eeal
In the evolution equations (\ref{KGreen}) and 
(\ref{Kspectral}) for the statistical and 
spectral function the Eq.\ (\ref{IRFR}) is replaced by
\bea
{\bf K}_{a}^F(x,y) \simeq
\frac{\lambda}{3N} \phi_b (x) F_{ba}(x,y) + 
\left( \frac{\lambda}{3N} \right)^2 \int_0^{x^0} dz\, 
\Pi_\rho(x,z)\phi_b(z)F_{ba}(z,y) 
&&\nonumber\\
-\left( \frac{\lambda}{3N} \right)^2 
\int_0^{y^0} dz\, \Pi_F(x,z) \phi_b(z)\rho_{ba}(z,y), 
&&\\
{\bf K}_{a}^{\rho}(x,y) \simeq  
\frac{\lambda}{3N} \phi_{b} (x) \rho_{ba}(x,y)
 + \left( \frac{\lambda}{3N} \right)^2 \int_{y^0}^{x^0} dz\,  
\Pi_\rho(x,z) \phi_b(z)\rho_{ba}(z,y),
&&\nonumber
\eea
and Eq.\ (\ref{DRFR}) simplifies considerably to
\bea
{\bf \hat D}_{ F}(x,y) &\simeq &  \frac{\lambda}{3 N}
\left[ \phi_a(x)F_{ab}(x,y)\phi_b(y) - \Pi_F(x,y)\right], 
\nonumber\\ 
{\bf \hat D}_{\rho}(x,y) &\simeq&  \frac{\lambda}{3 N}  
\left[ \phi_a(x)\rho_{ab}(x,y)\phi_b (y) - \Pi_{\rho}(x,y)\right].
\eea

\section{Summary}

We have derived the $2PI$ effective action $\Gamma[\phi,G]$ for the 
$O(N)$ model using the $2PI$--$1/N$ expansion to next-to-leading order.
A detailed discussion of the classification of diagrams was presented. 
The equations of motion for the field expectation value $\phi$ and the
two-point function $G$ were calculated without further approximations.   
We showed the equivalence of the direct calculation with the
auxiliary-field formulation. 

A detailed, but separate investigation would be necessary in order to
discuss the question of the nonperturbative renormalizability of
$\Gamma[\phi,G]$ and the evolution equations derived above within the
approximations considered in this paper. In principle, this problem may
be treated following methods outlined in Ref.\ \cite{vanHees:2002js}.
However, concerning the applications we have in mind and which are listed
in the introduction, we emphasize that the physics of these problems is
dominated by soft excitations and requires a finite cutoff. Therefore,
from the practical point of view the important next step is to solve these
equations along the lines of Refs.\ \cite{Berges:2001fi,Aarts:2001yn}
using a straightforward lattice discretization.

\vspace{0.5cm}

\noindent
{\bf Acknowledgements}\\
J.~B.\ thanks J{\"o}rn Knoll and Hendrik van Hees for interesting discussions.
G.~A.\ was supported by the Ohio State University through a Postdoctoral 
Fellowship and by the U.~S.\ Department of Energy under Contract No.\ 
DE-FG02-01ER41190.
D.~A.\ is supported by DFG, project FOR 339/2-1.

\appendix

\section{}
\label{modifiedcounting}

In this section we discuss possible further approximations consistent with
an expansion in powers of $1/N$ and comment on some aspects of Goldstone's
theorem.  We note that within the NLO approximation the replacement
\begin{equation}
\tr\, G^2 \to (\tr\, G)^2 / N
\label{replacement}
\end{equation}
in the expressions for the diagrams discussed in Sec.\ \ref{Nexpansion} is
correct up to higher order terms. This corresponds to the replacement
$G_{ab}(x,y)G_{ab}(x,y)$ $\to$ $[G_{aa}(x,y)]^2/N$ in the functions ${\bf
B}(G)$ of Eq.\ (\ref{Feq}) and ${\bf I}(G)$ of Eq.\ (\ref{ieqab}). One
observes that the resulting expressions can no longer be represented by
the diagrams of Sec.\ \ref{Nexpansion}. To verify this replacement up to
NNLO corrections we note that (for given space-time coordinates) $G$ can be
diagonalized by virtue of $O(N)$ rotations. In particular, $G$ is diagonal
up to subleading corrections. This can be seen explicitly from the LO
solution of Eq.\ (\ref{stationary}) for the propagator, $G^{\rm (LO)}_{ab}
\sim \delta_{ab}$, which follows from the fact that the LO diagrams depend
only on the invariants $\tr\, G$ and $\phi^2$ (cf.\ Sec.\
\ref{Nexpansion}). Since the invariant $\tr\, G^2$ does not appear in LO
diagrams, the replacement (\ref{replacement}) is correct within the NLO
approximation. We stress here that (\ref{replacement}) has not been used
in the derivation of any equation presented in this
paper.

A similar argument cannot be applied to the invariant $\tr(\phi \phi G) =
\phi_a G_{ab} \phi_b$. To see this, it is sufficient to restrict our
attention to constant field configurations. In this case, it follows from
$O(N)$ symmetry that the most general form of the propagator is
\beq
 G_{ab}^{\rm (stat)} (\phi) = G_L(\phi^2) P_{ab}^L + 
 G_T(\phi^2) P_{ab}^T,
\label{Gconstphi}
\eeq   
where $P_{ab}^L = \phi_a \phi_b/\phi^2$ and $P_{ab}^T = \delta_{ab} -
P_{ab}^L$ are respectively the longitudinal and transverse projectors with
respect to the field direction. Using this decomposition, we first check
that the replacement (\ref{replacement}) is valid at NLO, in agreement
with the above general discussion. Indeed the difference
\beq
 \tr\, G^2 - \frac{(\tr\, G)^2}{N} = 
 (G_L - G_T)^2 \left( 1 - \frac{1}{N} \right)
 \sim N^0
\eeq
is subleading (recall that $\tr\, G^2 \sim \tr\, G \sim N$). However,
\beq
 \tr ( \phi \phi G ) - \frac{\phi^2 \tr\, G}{N} = 
 \phi^2 (G_L - G_T) \left( 1 - \frac{1}{N} \right)
 \sim N
\eeq
demonstrates that $\tr(\phi\phi G)$ cannot be replaced by $\phi^2\tr\, 
G/N$ up to higher order corrections.

In the remainder of this Appendix we want to show that Goldstone's theorem
is fulfilled at any order in the $2PI$--$1/N$ expansion. Following Sec.\
\ref{Nexpansion} the $2PI$ effective action can be written as a function
of the $O(N)$ invariants (\ref{oninvariants}) only,
\beq
 \Gamma [\phi,G] \equiv 
 \Gamma \left[ \phi^2, \tr (G^n), \tr (\phi \phi G^p) \right].
\label{Gammainv}
\eeq 
In the case of spontaneous symmetry breaking one has a constant $\phi \not
= 0$ and the propagator can be parametrized as in Eq.\ (\ref{Gconstphi}).
The standard $1PI$ effective action $\Gamma_{1PI}[\phi]$ is obtained by
evaluating the $2PI$ effective action at the stationary value
(\ref{stationary})  for $G$ \cite{Cornwall:1974vz}, and the mass matrix
${\cal M}_{ab}$ can then be obtained from
\beq  
{\cal M}_{ab} \sim \left. \frac{\delta^2 
\Gamma[\phi,G^{\rm (stat)} (\phi)]}{\delta
\phi_a\delta \phi_b}\right|_{\phi=\rm const}.
\eeq
If $\Gamma_{1PI}[\phi]$ is calculated from (\ref{Gammainv}) and 
(\ref{Gconstphi}) one observes that indeed the $1PI$ effective action  
depends only on one invariant, $\phi^2$. The form of the mass matrix 
${\cal M}_{ab}$ can now be inferred straightforwardly from 
$\Gamma_{1PI}[\phi]$. To obtain the effective potential $U(\phi^2/2)$, we 
write
\beq
 \left. \Gamma_{1PI} [\phi] \right|_{\phi=\rm const} 
= \Omega_{d+1} U (\phi^2/2),
\eeq
where $\Omega_{d+1}$ is the $d+1$ dimensional euclidean volume. The 
expectation value of the field is given by the solution of the 
stationarity equation (\ref{phistationary}) which becomes
\beq
\label{minphi}
 \frac{\partial U(\phi^2/2)}{\partial \phi_a}  = 
 \phi_a \, U^\prime = 0,
\eeq
where $U^\prime \equiv \partial U/ \partial(\phi^2/2)$ and similarly for 
higher derivatives. The mass matrix reads  
\beq
 {\cal M}^2_{ab} =
 \frac{\partial^2 U(\phi^2/2)}{\partial \phi_a \partial \phi_b} 
 =  \delta_{ab} U^\prime + 
 \phi_a \phi_b U^{\prime \prime} = 
 ( U^\prime + \phi^2 U^{\prime \prime} ) P_{ab}^L + U^\prime P_{ab}^T.
\eeq
In the symmetric phase ($\phi_a=0$) one finds that all modes have equal
mass squared ${\cal M}^2_{ab} = U^\prime\delta_{ab}$.  In the broken
phase, with $\phi_a \neq 0$, Eq.~(\ref{minphi}) implies that the mass of
the transverse modes $\sim U^\prime$ vanishes identically in agreement
with Goldstone's theorem. For a similar discussion, see Ref.\ 
\cite{Verschelde:2000ta}.  
Truncations of the $2PI$ effective action 
may not show manifestly the presence of massless
transverse modes if one considers the solution $G^{(\rm stat)}$ of Eq.\
(\ref{stationary}) instead of the second variation of $\Gamma[\phi,G^{(\rm
stat)}]$ for constant fields.  For an early discussion of this point see
Ref.~\cite{Baym:qb} as well as the comments in Ref.\ \cite{vanHees:2002js}.

\section{}
\label{AppB}

In this Appendix we present the equations for the statistical and spectral 
functions preserving the nested integral structure and keeping the ``chain 
of bubbles'' ${\bf I}(x,y;G)$ as the basic quantity. 

All local contributions can be combined in an effective mass parameter 
\beq
M_{ab}^2(x) = [m^2+\bar\chi(x)]\delta_{ab} + 
\frac{\lambda}{3N}\left[ \phi_a(x)\phi_b(x) + G_{ab}(x,x)\right],
\eeq
and Eq.\ (\ref{geeq}) can be written as
\beq
\label{eqB2}
-\left[ \square_x\delta_{ac} + M_{ac}^2(x)\right] G_{cb}(x,y) = 
i\delta_{ab}\delta_\C(x-y) + i\int_z \Sigma_{ac}(x,z)G_{cb}(z,y),
\eeq
with the ``nonlocal'' self-energy at NLO (we suppress the $G$ dependence)
\beq
\Sigma_{ab}(x,y) = -\frac{\lambda}{3N}\left\{ {\bf 
I}(x,y)\left[\phi_a(x)\phi_b(y) + G_{ab}(x,y)\right] + {\bf 
P}(x,y) G_{ab}(x,y)\right\}.
\eeq
Here we defined
\bea
{\bf P}(x,y) &=& -\frac{\lambda}{3N}\int_{uv} {\bf B}^{-1}(x,u)\Delta(u,v) 
{\bf B}^{-1}(v,y),\\
\Delta(x,y) &=& -\phi_a(x)G_{ab}(x,y)\phi_b(y).
\eea
Eq.\ (\ref{eqB2}) results in the standard time evolution equations for $F$ 
and $\rho$ \cite{Aarts:2001qa}
\bea 
\left[ \square_x \delta_{ac} + M_{ac}^2(x) \right] F_{cb}(x,y) = 
- \int_0^{x^0} dz\, \Sigma^{\rho}_{ac}(x,z) F_{cb}(z,y)
\nonumber && \\ 
+ \int_0^{y^0} dz\, \Sigma^{F}_{ac}(x,z)\rho_{cb}(z,y),&&\\
\left[\square_x \delta_{ac} + M_{ac}^2(x) \right] \rho_{cb} (x,y) = 
 -\int_{y^0}^{x^0} dz\,  \Sigma^{\rho}_{ac} (x,z)\rho_{cb} (z,y), &&
\eea
with
\beq
M_{ab}^2(x) = [m^2+\bar\chi(x)]\delta_{ab} + 
\frac{\lambda}{3N}\left[ \phi_a(x)\phi_b(x) + F_{ab}(x,x)\right],
\eeq
and the nonlocal self-energies 
\bea
\Sigma^{F}_{ab}(x,y) & = &  - \frac{\lambda}{3 N}\Big\{
 {\bf I}_F(x,y)\left[ \phi_a (x)\phi_b(y) + F_{ab}(x,y) \right] -
 \frac{1}{4} {\bf I}_{\rho}(x,y)\rho_{ab}(x,y)  
 \nonumber\\
 && \qquad \quad + {\bf P}_F(x,y)F_{ab}(x,y) -
 \frac{1}{4} {\bf P}_{\rho}(x,y)\rho_{ab} (x,y) \Big\},
 \label{ASFFR}\\
\Sigma^{\rho}_{ab} (x,y) & = &  - \frac{\lambda}{3 N}\Big\{ 
 {\bf I}_{\rho}(x,y) \left[ \phi_a (x)\phi_b(y) + F_{ab} (x,y) \right] 
 + {\bf I}_{F}(x,y)\rho_{ab} (x,y)  
 \nonumber\\
 && \qquad \quad + {\bf P}_{\rho}(x,y) F_{ab}(x,y) + 
 {\bf P}_{F}(x,y) \rho_{ab} (x,y) \Big\}. 
 \label{ASRFR}
\eea
The functions ${\bf I}_{F}$ and ${\bf I}_{\rho}$ satisfy 
\cite{Berges:2001fi}
\bea
{\bf I}_{F}(x,y) =  -\frac{\lambda}{3N} 
 \Pi_F (x,y) +   \frac{\lambda}{3N}\int_{0}^{x^0} dz\,
 {\bf I}_{\rho}(x,z)\Pi_F (z,y) &&  
\nonumber\\
 - \frac{\lambda}{3N}\,\int_0^{y^0} dz\,  
 {\bf I}_F(x,z) \Pi_{\rho} (z,y), &&
 \nonumber\\
{\bf I}_{\rho}(x,y) = - \frac{\lambda}{3N}\Pi_{\rho} (x,y)
 + \frac{\lambda}{3N} \int_{y^0}^{x^0} dz\,  
 {\bf I}_{\rho}(x,z) \Pi_{\rho} (z,y),
 &&\label{AIRFR}
\eea
and the nested integrals are
\bea 
 {\bf P}_{F} (x,y) &=& - \frac{\lambda}{3N} \Bigg\{ \Delta_F (x,y)
 \nonumber\\
 && - \int_0^{x^0} dz\, \left[ \Delta_{\rho} (x,z){\bf I}_F (z,y) +
 {\bf I}_{\rho} (x,z)\Delta_F (z,y) \right] 
 \nonumber\\
 && + \int_0^{y^0} dz\, \left[  \Delta_F (x,z) {\bf I}_{\rho} (z,y) +
 {\bf I}_F (x,z) \Delta_{\rho} (z,y)\right] 
 \nonumber\\
 && - \int_0^{x^0} dz\, \int_0^{y^0} dv\,
 {\bf I}_{\rho} (x,z) \Delta_F (z,v) {\bf I}_{\rho} (v,y)
 \nonumber\\
 && + \int_0^{x^0} dz\, \int_0^{z^0} dv\,
 {\bf I}_{\rho} (x,z) \Delta_{\rho} (z,v) {\bf I}_F (v,y)
 \nonumber\\
 && + \int_0^{y^0} dz\, \int_{z^0}^{y^0} dv\,
 {\bf I}_F (x,z) \Delta_{\rho} (z,v) {\bf I}_{\rho} (v,y) \Bigg\},
\label{eqB11}
\eea
and 
\bea
 {\bf P}_{\rho} (x,y) & = & - \frac{\lambda}{3N} \Bigg\{ 
\Delta_{\rho} (x,y)
 \nonumber\\
 && - \int_{y^0}^{x^0} dz\, 
 \left[ \Delta_{\rho} (x,z) {\bf I}_{\rho} (z,y) + 
 {\bf I}_{\rho} (x,z) \Delta_{\rho} (z,y) \right] 
 \nonumber\\
 && +  \int_{y^0}^{x^0} dz\, \int_{y^0}^{z^0} dv\,
 {\bf I}_{\rho} (x,z) \Delta_{\rho} (z,v) {\bf I}_{\rho} (v,y) \Bigg\},
\label{eqB12}
\eea
with $\Delta_F(x,y) = -\phi_a(x) F_{ab}(x,y) \phi_b(y)$ and 
$\Delta_\rho(x,y) = -\phi_a(x) \rho_{ab}(x,y) \phi_b(y)$. 
The RHS in Eq.\ (\ref{Rphi}) for the field expectation value reads  
\bea
\nonumber
&&{\bf K}^F_a(x,x) = \frac{\lambda}{3N} F_{ab}(x,x)\phi_b(x) \\ 
&& \quad  -
 \frac{\lambda}{3N} \int_0^{x^0} dy \, 
 \left[ {\bf I}_{\rho} (x,y) F_{ab} (x,y) + 
 {\bf I}_F (x,y) \rho_{ab} (x,y) \right] \phi_b (y).
\eea
Note that the nested time integrals in Eqs.\ (\ref{eqB11}, \ref{eqB12})  
have been eliminated in the equations discussed in Sects.\ \ref{Secteom},
\ref{secevolution} by a convenient choice of auxiliary variables.

\end{document}